\documentclass[pra,twocolumn,showpacs,amsmath,amssymb]{revtex4-1}

\usepackage{graphicx}
\usepackage{amssymb}
\usepackage{amsmath}
\usepackage{dcolumn}
\usepackage{bm}
\usepackage{mathrsfs}
\usepackage{color}

\begin{document}

\title{Input-output theory of the unconventional photon blockade}

\author{H. Flayac}
\affiliation{Institute of Theoretical Physics, \'{E}cole Polytechnique F\'{e}d\'{e}rale de Lausanne EPFL, CH-1015 Lausanne, Switzerland}
\author{V. Savona}
\affiliation{Institute of Theoretical Physics, \'{E}cole Polytechnique F\'{e}d\'{e}rale de Lausanne EPFL, CH-1015 Lausanne, Switzerland}

\begin{abstract}
We study the unconventional photon blockade, recently proposed for a coupled-cavity system, in presence of input and output quantum fields. Mixing of the input or output channels still allows strong photon antibunching of the output field, but for optimal values of the system parameters that differ substantially from those that maximize antibunching of the intracavity field. This result shows that the specific input-output geometry in a photonic system determines the optimal design in view of single-photon device operation. We provide a compact analytical formula that allows finding the optimal parameters for each specific system geometry.
\end{abstract}
\pacs{42.50.-p,03.65.-w,71.36.+c}
\maketitle

\section{Introduction}

The generation of single photons is a crucial requirement in information and communication technology \cite{Kimble2008}. A single-photon source typically relies on a system able of producing subpoissonian light when driven by a classical light field. This mechanism requires a strong optical nonlinearity, such that the optical response to one photon can be modulated by the presence or absence of a single photon in the system -- the so-called photon blockade. Combined with the requirement of miniaturization for integrability and scalability purposes, a strong nonlinearity is typically achieved by increasing the time duration of the interaction between light and a small nonlinear system (e.g. a two-level optical transition), by means of a resonant optical cavity. This basic paradigm -- from which the research area called cavity quantum electrodynamics \cite{Mabuchi2002} stems -- was proposed long ago \cite{Carmichael1991}, and has been meanwhile experimentally demonstrated in atomic \cite{Birnbaum2005}, semiconductor \cite{He2013,Matthiesen2012,Reinhard2012}, and superconducting \cite{Bozyigit2011} hybrid systems, while theoretical proposals have been formulated for optomechanical systems \cite{Ludwig2012,Nunnenkamp2011,Rabl2011,Stannigel2012}. Yet, to achieve a sizeable photon blockade in these systems, the energy scale characterizing the optical nonlinearity must be large, i.e. it must exceed the optical losses, thus posing a severe technological challenge.

Recently, a new paradigm for the generation of subpoissonian light was proposed \cite{LiewPRL2010}. Such {\em unconventional photon blockade} (UPB) \cite{Carusotto2013} differs from the conventional mechanism in that the blockade is enforced by quantum interference between multiple excitation pathways \cite{BambaPRA2011}, rather than by an effective photon-photon repulsion induced by the strong nonlinearity. The UPB mechanism occurs in a system of two coupled cavities, where the linear coupling strength is the dominant energy scale, while a very weak third-order nonlinearity characterizes the two resonators. UPB displays a strong resonant behavior, as a function of the two resonant energies and the coupling strength. It is essentially thanks to this resonant character that the effect of a very weak nonlinearity can be amplified almost at will, to produce photon blockade. UPB holds great promise as an alternative paradigm in view of an integrated and scalable technology for single-photon generation \cite{FerrettiNJoP2013}. Several possible implementations of this paradigm are currently being considered, for example in polaritonic \cite{Bamba2011}, optomechanical \cite{Savona2013,Xu2013}, or photonic crystal systems \cite{Majumdar2012,FerrettiNJoP2013}.

Each of these possible implementations is based on a system design that includes a specific scheme of input and output channels. In the original proposal, the basic UPB mechanism occurs only for the intracavity field of one of the two cavities. In a realistic implementation, each of the two cavities couples predominantly to a different input and output channel, but unavoidably some mixing between the two input or the two output channels must be expected. An example could be that of a system based on a photonic crystal slab \cite{FerrettiNJoP2013}, in which the mixing is simply produced by proximity between the cavities and the input-output channels. Given the interferential nature of UPB, a very natural expectation is that such a mixing may affect the mutual phase relation between fields, ultimately suppressing the antibunching \cite{Bamba2011,Majumdar2012}.

Here, we study the UPB mechanism by modeling quantum input and output channels with arbitrary degree of mixing. We compute the two-photon correlation function at zero delay, both numerically by fully solving the system master equations, and analytically in the limit of weak driving field. We demonstrate that, contrarily to common expectations, an optimal condition for UPB still exists for arbitrary degree of mixing, but it occurs for system parameters -- particularly the resonant frequencies of the two cavities -- that differ considerably from those derived for the intracavity field \cite{BambaPRA2011}. Our result thus shows that each system-specific design of input-output channels determines different optimal parameters for UPB. These parameters must then be modeled appropriately {\em before fabrication}, as they determine the optimal system design. Here, we provide a compact analytical tool that allows to easily link the input-output mixing rates to the optimal design of the two cavities.

\section{Intracavity fields}

In cavity quantum electrodynamics, several specific systems display nonlinear optical properties that can be mapped -- under appropriate conditions -- onto the simple model of an oscillator with a third order nonlinearity. We recall here the three most common cases. First, an optical cavity embedding a Kerr optical medium \cite{Ferretti2012}. Second, an optical cavity whose resonant mode is coupled to the optical transition of a two-level system, well described by the Jaynes-Cummings model which has widespread applications to atomic \cite{Birnbaum2005}, semiconductor \cite{Reinhard2012}, and superconducting systems \cite{Lang2011}. In this case, the equivalence holds only in the limit of large detuning between the cavity mode and the two-level transition, compared to the coupling strength \cite{Boissonneault2009,Koch2009}. Third, an optomechanical system in which the optical density inside the cavity is coupled to the displacement of a mechanical oscillator \cite{Aldana2013,Rabl2011}. Given this broad range of currently investigated systems, it is therefore interesting to study quantum optical effects using the model of an oscillator with third order nonlinearity.

Here, we consider the system of two coupled single-mode cavities, described respectively by Bose operators $\hat{a}_{1,2}$ and $\hat{a}^\dagger_{1,2}$, as originally studied in Ref.\cite{LiewPRL2010}. Both modes are characterized by a weak third-order nonlinearity of strength $U$ and are driven by continuous-wave fields having the same frequency $\omega_L$ and distinct amplitudes $F_{1,2}$. In the frame rotating at $\omega_L$, the intracavity Hamiltonian reads:
\begin{eqnarray}
\label{Hamiltonian}
\nonumber \cal H &=& \sum\limits_{j = 1,2} \left[{{E_j}\hat{a}_j^\dag {\hat{a}_j} + {U}\hat{a}_j^{\dag 2}\hat{a}_j^2 + {F_j}\hat{a}_j^\dag  + F_j^*{\hat{a}_j}}\right]\\
&-& J\left( {\hat{a}_1^\dag {\hat{a}_2} + \hat{a}_2^\dag {\hat{a}_1}} \right)\,,
\end{eqnarray}
where $E_{1,2}$ are the mode energies expressed with respect to $\omega_L$.
The system dynamics is governed by the following quantum master equation for the density matrix $\hat\rho$
\begin{equation}
\label{rhoDyn}
i\frac{{d\hat\rho }}{{dt}} = \left[ {\cal H,\hat{\rho} } \right] + \mathcal{L}^{\left(\rm{rad}\right)}+\mathcal{L}^{\left(\rm{pd}\right)}\,,
\end{equation}
where
\begin{eqnarray}
\label{Lrad}
\mathcal{L}^{\left(\rm{rad}\right)}&=&{\frac{i}{2}\sum\limits_{j = 1,2}  \Gamma_j\left( {2{\hat{a}_j}\hat{\rho} \hat{a}_j^\dag  - \left\{ {\hat{a}_j^\dag {\hat{a}_j},\hat{\rho} } \right\}} \right)}\\
\label{Lpd}
{{\cal L}^{\left(\rm{pd}\right)}} &=& \frac{{i}}{2}\sum\limits_{j = 1,2} {\Gamma_j^{\left( {\rm{pd}} \right)}} \left( {2\hat{a}_j^\dag {\hat{a}_j}\hat{\rho} \hat{a}_j^\dag {\hat{a}_j} - \left\{ {\hat{a}_j^\dag {\hat{a}_j}\hat{a}_j^\dag {\hat{a}_j}},\hat{\rho} \right\}} \right)
\end{eqnarray}
are Lindblad superoperators modeling, in the Markov limit, respectively radiative losses at rates $\Gamma_{1,2}$ and pure dephasing processes at rates $\Gamma_{1,2}^{\left(\rm{pd}\right)}$. Below, we will solve numerically Eq. (\ref{rhoDyn}) in the stationary limit $d\hat{\rho}/dt=0$, within a truncated Hilbert space \cite{LiewPRL2010}.

\begin{figure}[ht]
\includegraphics[width=0.45\textwidth,clip]{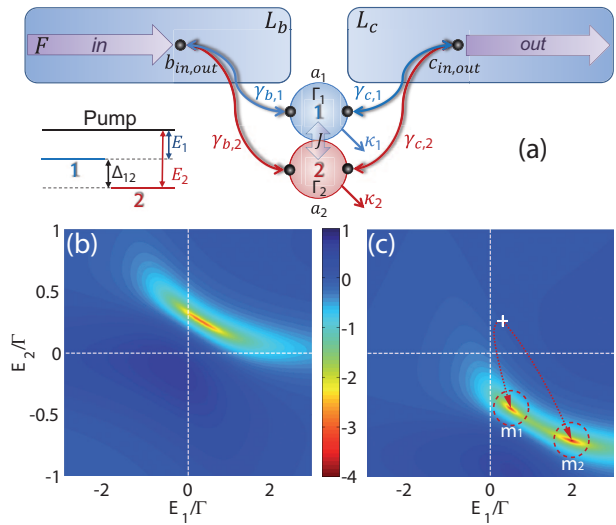}\\
\caption{(a) Sketch of the system. Two coupled cavities are evanescently linked to two semi-infinite waveguides $L_b$ and $L_c$. The blue and red double-arrows denote input-output couplings. The inset shows the energy levels and associated detunings. Lower panels: $\log_{10}[g_{\rm{out}}^{\left( 2 \right)}(0)]$ as a function of $E_1$ and $E_2$ for $\gamma_{1}=0.4\Gamma$ , (b) $\gamma_{2}=0$ and (c) $\gamma_{2}=0.025\gamma_1$. The other parameters are $\kappa_{1,2}=\Gamma-2\gamma_{1,2}$, $U=0.012/\Gamma$, $J=0.5/\Gamma$ and $F=0.01/\Gamma$. The white cross marks the location of the global minimum of panel (b), while the circles indicate the two local minima, and the arrows highlight their displacement as $\gamma_2$ is increased.}
\label{Fig1}
\end{figure}

\begin{figure*}[ht]
\includegraphics[width=0.9\textwidth,clip]{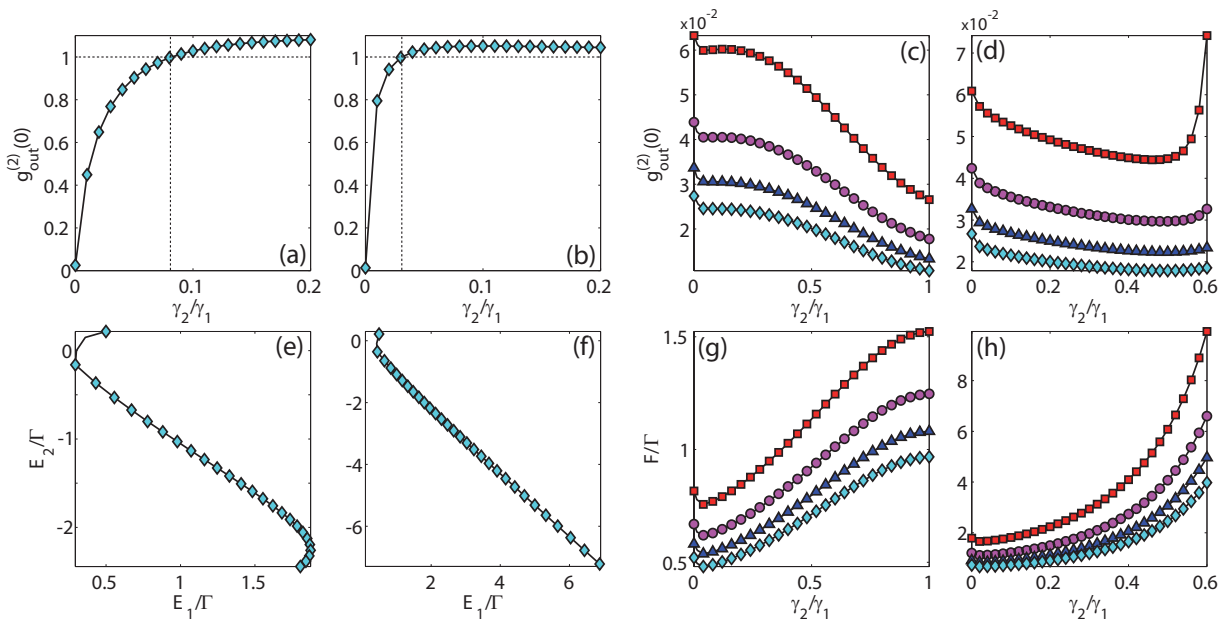}\\
\caption{Steady-state solutions of Eq.(\ref{rhoDyn}). (a): $g_{\rm{out}}^{\left( 2 \right)}(0)$ as a function of $\gamma_2/\gamma_1$ at constant detunings $E_{1,2}=\pm \Gamma /2\sqrt 3$ (white cross in Fig\ref{Fig1}(c)). (b): same as (a), when also accounting for the input mixing Eq.(\ref{F12}). These values do not depend significantly on $\gamma_1$. (c) and (d): Same as respectively (a) and (b), but tracking $\min(g_{out}^{(2)}(0))$ as a function of $E_1$ and $E_2$, for each value of the abscissa. Red squares, purple disks, blue triangles and light-blue diamonds correspond to $\gamma_1=0.2, 0.3, 0.4, 0.5$ respectively. The corresponding displacement on the $(E_1,E_2)$-plane is shown in panels (e) and (f) respectively. This displacement is essentially independent of $\gamma_1$. (g) and (h): Pump amplitude required to give a constant occupation $N_{\rm{out}}=10^{-3}$ for the data in panels (c) and (d) respectively.}
\label{Fig2}
\end{figure*}

\section{Input-Output fields}
We consider input and output lines as sketched in Fig.\ref{Fig1}(a). We assume two semi-infinite waveguides $L_b$ and $L_c$ evanescently coupled to the cavities, in the wake of the proposal of Ref.\cite{FerrettiNJoP2013}. Each waveguide acts simultaneously as an input and an output channel for the two-cavity system. The corresponding Bose operators for the input and output modes are denoted as $\hat{b}_{\rm{in}}$, $\hat{b}_{\rm{out}}$, $\hat{c}_{\rm{in}}$ and $\hat{c}_{\rm{out}}$. The evanescent coupling of cavities 1 and 2 to waveguides $L_b$ and $L_c$ is quantified by the rates $\gamma_{b,1,2}$ and $\gamma_{c,1,2}$, respectively. The coherent driving field is conveyed through the waveguide $L_b$, while the device output is collected from the $L_c$ channel.

According to the input-output formalism of Collett and Gardiner \cite{CollettPRA1984,GardinerPRA1985}, the input, output and intracavity fields are linked through the boundary condition
\begin{eqnarray}
\label{bout}
\hat{b}_{{\rm{out}}}^{(\dag )}\left( t \right) &=& \hat{b}_{{\rm{in}}}^{(\dag )} + \sqrt {{\gamma _{b,1}}} \hat{a}_1^{(\dag )} + \sqrt {{\gamma _{b,2}}} \hat{a}_2^{(\dag )}\,,\\
\label{cout}
\hat{c}_{{\rm{out}}}^{(\dag )}\left( t \right) &=& \hat{c}_{{\rm{in}}}^{(\dag )} + \sqrt {{\gamma _{c,1}}} \hat{a}_1^{(\dag )} + \sqrt {{\gamma _{c,2}}} \hat{a}_2^{(\dag )}\,.
\end{eqnarray}
The rates $\gamma_{b,1,2}$ and $\gamma_{c,1,2}$ contribute, together with the intrinsic loss rate $\kappa_{1,2}$ of each cavity, to the total loss rates of the two cavity modes, i.e. $\kappa_{1,2}+\gamma_{b,1,2}+\gamma_{c,1,2}=\Gamma_{1,2}$. All these loss rates depend on the specific system type and cavity design. Here, in order to comply to the assumption made in Ref.\cite{LiewPRL2010}, we set $\Gamma_1=\Gamma_2=\Gamma$. Notice that this assumption generally leads to different intrinsic loss rates $\kappa_{1,2}$ for the two cavities. The extension to more general assumptions is however straightforward. Additionally we shall consider a system that is symmetric with respect to the input-output guides imposing $\gamma_{1,2}=\gamma_{b,1,2}=\gamma_{c,1,2}$. The ideal case described of Ref.\cite{LiewPRL2010} is then recovered by setting $\gamma_{2}=0$. A finite value of $\gamma_2$ expresses instead the fact that any output observable includes contributions from both intra-cavity fields. This is unavoidable in most experimental setups, where both the coupling between cavities 1 and 2 and the coupling to input-output lines are realized through spatial proximity, as suggested by the sketch in Fig.\ref{Fig1}(a).

For an arbitrary state of the input modes, correlations of the output fields would depend on cross-correlations between the input and intra-cavity fields, which in turn would require to model the input fields together with the system dynamics. However, if we assume only classical driving fields added to the quantum vacuum of the input-output channels, then all normally ordered cross-correlations between intra-cavity and input modes vanish, and correlations in the output channels can be expressed as functions of intra-cavity correlations only. Within this assumption, the average number of photons collected through $L_c$ reads
\begin{eqnarray}
\nonumber N_{\rm{out}}=\left\langle {\hat{c}^{\dag }_{\rm{out}}\hat{c}_{\rm{out}}} \right\rangle &=& {\gamma _{1}}\left\langle {\hat{a}_1^\dag {\hat{a}_1}} \right\rangle  + {\gamma _{2}}\left\langle {\hat{a}_2^\dag {\hat{a}_2}} \right\rangle\\
&+& 2\sqrt {{\gamma _{1}}{\gamma _{2}}} \left\langle {\hat{a}_1^\dag {\hat{a}_2}+\hat{a}_2^\dag {\hat{a}_1}} \right\rangle\,.
\end{eqnarray}
Similarly, the second order correlation function of the output field at zero delay is given by
\begin{eqnarray}
\label{g2out}
g^{\left( 2 \right)}_{\rm{out}}\left( 0 \right) &=& \frac{{\left\langle {\hat{c}^{\dag }_{\rm{out}} \hat{c}^{\dag }_{\rm{out}} \hat{c}_{\rm{out}}\hat{c}_{\rm{out}}} \right\rangle }}{N_{\rm{out}}^2}\,,\\
&=&\sum\limits_{j,k,l,m = 1,2} {\sqrt {{\gamma _{j}}{\gamma _{k}}{\gamma _{l}}{\gamma _{m}}} } \frac{{\left\langle {\hat a_j^\dag \hat a_k^\dag {{\hat a}_l}{{\hat a}_m}} \right\rangle }}{{N_{out}^2}}\,.
\end{eqnarray}

If the driving fields are delivered through the two waveguides, then the same mixing weights must hold given the system symmetry, namely
\begin{eqnarray}
{F_{1,2}} &=& \sqrt {\frac{\gamma_{1,2}}{\gamma_{1}+\gamma_{2}}} F\,,
\label{F12}
\end{eqnarray}
so that ${\left| F  \right|^2} = {\left| {{F_1}} \right|^2} + {\left| {{F_2}} \right|^2}$.

\section{Results}
The antibunching of the intracavity field is maximized for optimal values \cite{BambaPRA2011}
\begin{eqnarray}
E_{1,2}&\simeq&  \pm \Gamma /2\sqrt 3\,,\label{Eopt}\\
U&\simeq&  2{\Gamma ^3}/3{J^2}\sqrt 3\,.\label{Uopt}
\end{eqnarray}
Here, we set $U$ to its optimal value (\ref{Uopt}), and let the detunings $E_1$ and $E_2$ vary. Fig.\ref{Fig1}(a) shows the computed value of $\log_{10}[g^{(2)}_{out}(0)]$, as a function of $E_1$ and $E_2$, in the case $\gamma_2=0$, namely without mixing. The result reproduces exactly that for the intracavity field in Ref.\cite{LiewPRL2010}. When instead a finite value $\gamma_2=0.025\gamma_1$ is set, the optimal antibunching moves to different values of $E_1$ and $E_2$, and the original minimum splits into two distinct minima. Indeed, if the optimal values (\ref{Eopt}) and (\ref{Uopt}) are set, the value of $g^{(2)}_{out}(0)$ rapidly increases as a function of $\gamma_2/\gamma_1$, both when assuming output mixing only (Fig.\ref{Fig2}(a)) and if a corresponding input mixing (\ref{F12}) is introduced (Fig.\ref{Fig2}(b)). These results suggest that the optimal values of $E_1$ and $E_2$ must be found independently for each value of $\gamma_2$. We note that the absolute value of $\gamma_1$ has practically no influence on this behaviour which is solely determined by the ratio $\gamma_2/\gamma_1$.

To illustrate how the optimal values of $E_1$ and $E_2$ depend on the mixing, we plot on the ($E_1$, $E_2$)-plane the position of the minimum labeled $m_1$ in Fig.\ref{Fig1}(c), for $\gamma_2/\gamma_1$ uniformly increasing from $0$ to $0.5$, both in the case of output mixing only (Fig.\ref{Fig2}(e)) and including input mixing [Fig.\ref{Fig2}(f)]. Again, these values scarcely depend on the absolute value of $\gamma_1$. Correspondingly, Fig.\ref{Fig2}(c) and (d) show the value of  $g_{\rm{out}}^{(2)}(0)$ computed at increasing $\gamma_2/\gamma_1$, while tracking the optimal values of $E_1$ and $E_2$ as a function of this parameter. Here, different symbols denote different values of $\gamma_1$, showing that this parameter affects the actual value of the two-photon correlation. For these plots, the overall pump amplitude $F$ was also adjusted for each value of $\gamma_2/\gamma_1$ in order to keep the average photon occupation in the output mode constant, $N_{\rm{out}}=10^{-3}$. The corresponding values of $F$ are plotted in Fig.\ref{Fig2}(g) and (h) respectively. These data show that the unconventional antibunching can indeed be preserved -- or even slightly improved -- when tuning the optimal values of $E_1$ and $E_2$ to the ratio of output coupling rates $\gamma_2/\gamma_1$ characterizing the specific system under investigation.

When considering a given engineered sample, the detuning between the two cavities $\Delta_{12}=E_1-E_2$ is generally determined by the specific sample design and fabrication. It is therefore relevant, in view of an experiment, to study the optimal value of the detuning $E_1$ for each given value $\Delta_{12}$. The corresponding results (accounting for the mixed input) are shown in Fig.\ref{Fig3}(a) and (b). Panel (a) shows the optimal value of $g_{\rm{out}}^{(2)}(0)$ computed as a function of $\Delta_{12}$ for three different values of $\gamma_2$, while panel (b) shows the corresponding optimal value of $E_1$ and the pump amplitude (inset) required to have $N_{\rm{out}}=10^{-3}$. As soon as the output fields are mixed we see that optimal antibunching requires $\Delta_{1,2}>0$, differently from the ideal case\cite{LiewPRL2010,BambaPRA2011} where $\Delta_{12}\simeq0$. In the cases where $\gamma_{2}\neq0$, two distinct minima in $g_{\rm{out}}^{(2)}$ appear, consistently with Fig.\ref{Fig1}(c).

Finally we have analyzed in Fig.\ref{Fig3}(c),(d) the impact of pure dephasing, as introduced in Eq. (\ref{Lpd}), for three values of $\Gamma^{(\rm{pd})}=\Gamma_{1}^{(\rm{pd})}=\Gamma_{2}^{(\rm{pd})}$. While the optimal values of $E_1$ and $E_2$ are only slightly affected [panel (d)], the corresponding optimal value of $\min(g_{\rm{out}}^{(2)}(0))$ increases significantly [panel(c)], as expected for the unconventional mechanism which strongly relies on quantum interference. In particular, antibunching starts being suppressed as soon as the pure dephasing rate becomes comparable to the nonlinear energy $U$. This is again expected, as the destructive quantum interference between different excitation pathways leading to the two-photon state is enforced by a nonlinear energy shift of magnitude $U$. It can be concluded that the sensitivity to pure dephasing is not dramatically modified by mixing in the input or in the output channels.

\begin{figure}[ht]
\includegraphics[width=0.45\textwidth,clip]{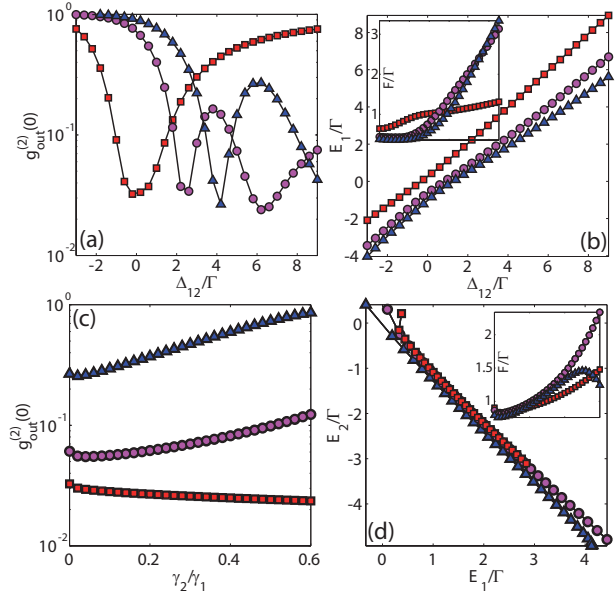}\\
\caption{(a) Minimal value of $g_{\rm{out}}^{(2)}(0)$ as a function of the detuning $\Delta_{12}$, obtained by varying $E_1$. (b) Corresponding optimal value of $E_1$. Inset: value of $F$ required to give a constant occupation $N_{\rm{out}}=10^{-3}$. For both plots, $\gamma_{1}=0.2\Gamma$ (red squares), $0.3\Gamma$ (purple disks) and $0.4\Gamma$ (blue triangles). (c) Impact of the pure dephasing on $\min(g_{\rm{out}}^{(2)}(0))$ for $\Gamma^{(\rm{pd})}=0$ (red squares), $0.01 U$ (purple disks) and $0.1 U$ (blue triangles). (d) Corresponding points on the $(E_1,E_2)$-plane. Inset: value of $F$ required to impose a constant occupation $N_{\rm{out}}=10^{-3}$.}
\label{Fig3}
\end{figure}

UPB relies on the fact that, thanks to destructive quantum interference between multiple excitation pathways leading to the state with two photons in the first cavity, the amplitude of this particular number-state in the stationary state of the system vanishes under appropriate choice of the system parameters. Then, in the limit of vanishing pump amplitude $F$, the probability of having $N_1>2$ is negligible and $g_{\rm{out}}^{(2)}(0)$ actually vanishes as a result. In presence of output mixing however, Eqs.(\ref{bout}) and (\ref{cout}) show that an analogous canceling of the two-photon amplitude in the output field now requires suppressing the two-photon occupation of the mode associated to the linear superposition $\sqrt{\gamma_1}\hat{a}_1+\sqrt{\gamma_2}\hat{a}_2$. This is made possible by a similar quantum interference scheme as in the ideal case, where however optimal conditions occur for different values of the detunings $E_1$ and $E_2$. This is ultimately the reason why output mixing does not actually suppress antibunching but rather moves it to a different optimal point in parameter space. More insight on this result can be obtained by carrying out an analysis to leading order in $F$, similar to the one presented for the ideal case in Ref.\cite{BambaPRA2011}. This leads to a compact analytical expression for $g_{\rm{out}}^{(2)}(0)$ in the limit $F\to0$, derived in the appendix, that can be easily adopted to optimize the parameters $E_1$ and $E_2$ in each specific case.

The main indication coming from the present study is that, in any attempt to experimentally design a coupled-cavity system for the detection of unconventional photon blockade, the input and output mixing must be accurately modeled before fabrication, in order to assess the corresponding optimal detuning $\Delta_{12}=E_1-E_2$ between the two cavities.

\section{Conclusion}
In summary, we have studied the unconventional photon blockade in the context of an input-output theory of the open quantum system, in order to assess how unavoidable mixing between the two input or the two output channels affects the photon antibunching. Our findings clearly show that the photon antibunching is not suppressed but rather just displaced in a different region of the system's parameter space. Unconventional photon blockade was recently proposed as a very effective mechanism to produce a strongly subpoissonian photon field in presence of arbitrarily weak nonlinearities. This mechanism holds great promise in view of the realization of integrated single-photon sources, that could operate even by only relying on the very weak background third-order nonlinearity of the dielectric material \cite{FerrettiNJoP2013}, and could be the mechanism of choice to observe quantum effects in hybrid systems -- such as i.e. optomechanical systems \cite{Savona2013,Xu2013} -- where strong single-photon nonlinearities are far from being achieved. It should be observed that, to operate as an on-demand single photon source, a system displaying the unconventional photon blockade would additionally require pulsed operation, for which the unconventional blockade mechanism is subject to some restrictions as discussed by Bamba {\em et al.} \cite{Bamba2011}. In particular, in order for antibunching to occur in the pulsed regime, the bandwidth of the pump pulse must be smaller than the cavity loss rate $\Gamma$, while its time duration must be shorter than the timescale of the quantum correlations, set by $J^{-1}$. These two conditions are barely met for the typical regime $J>\Gamma$, in which the unconventional blockade occurs. A possible way to overcome this limitation might be to filter the output either in frequency or in time -- a task for which the present input-output theory is the appropriate tool. Spectral filtering in particular has been shown to enhance photonic antibunching of the conventional type \cite{Qu1992}. Alternatively, an appropriate shaping of the pump pulse might also improve single-photon operation. These ideas however require verification through a time-resolved analysis of the unconventional photon blockade, which will be the object of future work.

Our study shows that, in order to produce the optimal conditions for the unconventional blockade, the system-specific input-output relations play a crucial role, affecting dramatically the optimal system design. The present result clearly indicates what are the optimal system parameters and provides a tool for their evaluation, that can be easily adopted in the several contexts -- ranging from photonic crystal \cite{Majumdar2012,FerrettiNJoP2013} to optomechanical \cite{Savona2013,Xu2013} or polaritonic \cite{LiewPRL2010,BambaPRA2011,Bamba2011} systems -- in which an experimental demonstration of the unconventional photon blockade is currently being sought.

\begin{acknowledgments}
We are grateful to D. Gerace and S. Savasta for fruitful discussions. Our work was supported by NCCR Quantum Photonics (NCCR QP), research instrument of the Swiss National Science Foundation (SNSF).
\end{acknowledgments}

\appendix*

\

\section{The weak pump limit}

In this section, we derive analytical expressions for the number of intracavity photons and the zero-delay two-photon correlation function, in the limit of vanishing driving field.

We start from the Hamiltonian (\ref{Hamiltonian}) and expand the intracavity field wavefunction on a Fock-state basis, truncated to the two-photon manifold as allowed by the assumption of weak driving field.
\begin{eqnarray}\label{psi}
\nonumber \left| \psi \right\rangle  &=& {C_{00}}\left| {00} \right\rangle + {C_{10}}\left| {10} \right\rangle + {C_{01}}\left| {01} \right\rangle\\
                                     &+& {C_{11}}\left| {11} \right\rangle + {C_{20}}\left| {20} \right\rangle + {C_{02}}\left| {02} \right\rangle
\end{eqnarray}
Here, $\left| jk \right\rangle=\left| j \right\rangle\otimes\left| k \right\rangle$ denotes a Fock state with $j$ photons in the first cavity and $k$ photons in the second one.
The steady state is found from the stationary solution of the nonlinear Schr\"{o}dinger equation $\tilde H\left|\psi\right\rangle=i\hbar{\partial _t}\left| \psi  \right\rangle$ written for the non-Hermitian Hamiltonian
\begin{equation}\label{HNH}
\tilde H = H - \frac{{i\Gamma }}{2}\sum\limits_{j = 1,2} {\hat{a}_j^\dag } {\hat{a}_j}
\end{equation}
with the further assumption $\Gamma_{1,2}=\Gamma$.
We assume equal coupling to the two input-output channels, namely $\gamma_{b,1,2}=\gamma_{c,1,2}=\gamma_{1,2}$ and obtain the following coupled set of equations for the coefficients $C_{jk}$
\begin{eqnarray}
\label{EqC1}
0 &=& {F_1}{C_{10}} + {F_2}{C_{01}}\\
\label{EqC2}
0 &=& {F_1}{C_{02}} + {F_2}\sqrt 2 {C_{11}}\\
\label{EqC3}
0 &=& {F_2}{C_{02}} + {F_1}\sqrt 2 {C_{11}}\\
\label{EqC4}
0 &=& {F_1}{C_{00}} + {\tilde E_1}{C_{10}} - J{C_{01}} + \underline {{F_1}\sqrt 2 {C_{20}}}  + \underline {{F_2}{C_{11}}} \\
\label{EqC5}
0 &=& {F_2}{C_{00}} + {\tilde E_2}{C_{01}} - J{C_{10}} + \underline {{F_2}\sqrt 2 {C_{02}}}  + \underline {{F_1}{C_{11}}} \\
\label{EqC6}
0 &=& {F_1}\sqrt 2 {C_{10}} + 2\left( {{{\tilde E}_1} + {U_1}} \right){C_{20}} - J\sqrt 2 {C_{11}}\\
\label{EqC7}
0 &=& {F_2}\sqrt 2 {C_{01}} + 2\left( {{{\tilde E}_2} + {U_2}} \right){C_{02}} - J\sqrt 2 {C_{11}}\\
\label{EqC8}
\nonumber 0 &=& {F_2}{C_{10}} + {F_1}{C_{01}} - J\sqrt 2 \left( {{C_{20}} + {C_{02}}} \right) + \left( {{{\tilde E}_1} + {{\tilde E}_2}} \right){C_{11}}\\
\end{eqnarray}
where $\tilde E_j=E_j-i\Gamma/2$. Here, the pump amplitudes are expressed as $F_j=\sqrt{\zeta_{j}}F$ where $\zeta_{1,2}=\gamma_{1,2}/(\gamma_{1}+\gamma_{2})$. To leading order, the coefficients $C_{jk}$ depend on the driving field amplitude as $C_{jk}\propto F^{j+k}$. Hence, we can eliminate the underlined terms in Eqs. (\ref{EqC4}) and (\ref{EqC5}), as they are of sub-leading order in $F$. By further imposing the normalization condition $C_{00}=1$, straightforward algebra leads to the following solutions
\begin{eqnarray}
\label{C10}
{C_{10}} &=& F\frac{{{\tilde E_2}\sqrt {{\gamma _{1}}}  + J\sqrt {{\gamma _{2}}} }}{{{J^2} - {{\tilde E}_1}{{\tilde E}_2}}}\\
\label{C01}
{C_{01}} &=& F\frac{{{\tilde E_1}\sqrt {{\gamma _{2}}}  + J\sqrt {{\gamma _{1}}} }}{{{J^2} - {{\tilde E}_1}{{\tilde E}_2}}}
\end{eqnarray}

\begin{widetext}
\begin{figure*}[t!]
\renewcommand{\figurename}{Fig. A}
\setcounter{figure}{0}
\includegraphics[width=1\textwidth,clip]{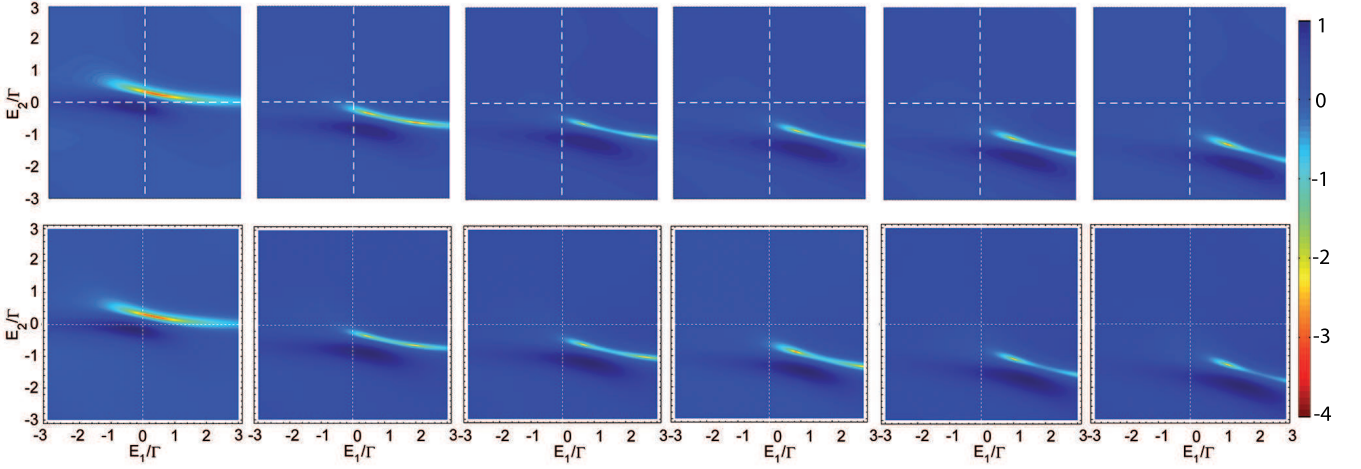}\\
\caption{Comparison of the numerical (upper panels) and analytical (lower panels) values of $\log_{10}(g^{(2)}_{\mathrm{out}}(0))$, as a function of $E_1$ and $E_2$, at fixed $\gamma_1=0.4\Gamma$, for $\gamma_2$ taking values from $\gamma_{2}=0$ to $\gamma_{2}=0.1\gamma_{1}$ by steps of $0.02 \gamma_1$ from left to right. Here $F=10^{-2}/\Gamma$, the other parameters are the same as for Fig.\ref{Fig1}.}
\label{FigS1}
\end{figure*}

\begin{eqnarray}
\label{C20}
{C_{20}} &=& {F^2}\frac{{\left[ {\tilde E_2^3 + \tilde E_2^2\left( {{{\tilde E}_1} + U} \right) + U{J^2}} \right]{\gamma _{1}} + {J^2}\left( {{{\tilde E}_1} + {{\tilde E}_2} + U} \right){\gamma _{2}} + 2J\sqrt {{\gamma _{1}}{\gamma _{2}}} \left( {{{\tilde E}_1} + {{\tilde E}_2}} \right)\left( {{{\tilde E}_2} + U} \right)}}{{\sqrt 2 \left( {{{\tilde E}_1}{{\tilde E}_2} - {J^2}} \right)\left( {\tilde E_1^2\left( {{{\tilde E}_2} + U} \right) + {{\tilde E}_1}\left[ {{{\left( {{{\tilde E}_2} + U} \right)}^2} - {J^2}} \right] + \tilde E_2^2U + {J^2}\left( {{{\tilde E}_2} - 2U} \right)} \right)}}\\
\label{C02}
{C_{02}} &=& {F^2}\frac{{\left[ {\tilde E_1^3 + \tilde E_1^2\left( {{{\tilde E}_2} + U} \right) + U{J^2}} \right]{\gamma _{2}} + {J^2}\left( {{{\tilde E}_1} + {{\tilde E}_2} + U} \right){\gamma _{1}} + 2J\sqrt {{\gamma _{1}}{\gamma _{2}}} \left( {{{\tilde E}_1} + {{\tilde E}_2}} \right)\left( {{{\tilde E}_1} + U} \right)}}{{\sqrt 2 \left( {{{\tilde E}_1}{{\tilde E}_2} - {J^2}} \right)\left( {\tilde E_2^2\left( {{{\tilde E}_1} + U} \right) + {{\tilde E}_2}\left[ {{{\left( {{{\tilde E}_1} + U} \right)}^2} - {J^2}} \right] + \tilde E_1^2U + {J^2}\left( {{{\tilde E}_1} - 2U} \right)} \right)}}\\
\nonumber {C_{11}} &=& {F^2}\frac{{J\left( {{{\tilde E}_1} + {{\tilde E}_2} + U} \right)\left[ {\left( {{{\tilde E}_2} + U} \right){\gamma _{1}} + J\left( {{{\tilde E}_1} + U} \right){\gamma _{2}}} \right]}}{{\left( {{{\tilde E}_1}{{\tilde E}_2} - {J^2}} \right)\left[ {\tilde E_1^2\left( {{{\tilde E}_2} + U} \right) + \tilde E_2^2\left( {{{\tilde E}_1} + U} \right) - {J^2}\left( {{{\tilde E}_1} + {{\tilde E}_2} - 2U} \right)} \right]}}\\
\label{C11}
&+& {F^2}\frac{{\sqrt {{\gamma _{1}}{\gamma _{2}}} \left[ {\left( {{{\tilde E}_1} + {{\tilde E}_2}} \right)\left( {{{\tilde E}_1}{{\tilde E}_2} + {J^2} + {U^2}} \right) + U\left( {\tilde E_1^2 + \tilde E_2^2} \right) + 2U{{\tilde E}_1}{{\tilde E}_2} + 2U{J^2}} \right]}}{{\left( {{{\tilde E}_1}{{\tilde E}_2} - {J^2}} \right)\left[ {\tilde E_1^2\left( {{{\tilde E}_2} + U} \right) + \tilde E_2^2\left( {{{\tilde E}_1} + U} \right) - {J^2}\left( {{{\tilde E}_1} + {{\tilde E}_2} - 2U} \right)} \right]}}
\end{eqnarray}
where we further assumed $U_{1,2}=U$. We then construct the system density matrix from $\rho=|\psi\rangle\langle\psi|$. By combining this result with the expressions for the input-output field operators, we obtain the following compact expression for the average photon occupation in the output field of the first cavity
\begin{equation}\label{Ninout}
{N_{\mathrm{out}}} = \left\langle {c_{\mathrm{out}}^\dag {c_{\mathrm{out}}}} \right\rangle = \mathrm{Tr}\left( {c_{\mathrm{out}}^\dag {c_{\mathrm{out}}}\rho } \right) = {\left| {F\frac{{\left( {{{\tilde E}_2}\sqrt {{\gamma _{1}}}  + J\sqrt {{\gamma _{2}}} } \right)\sqrt {{\gamma _{1}}}  + \left( {{{\tilde E}_1}\sqrt {{\gamma _{2}}}  + J\sqrt {{\gamma _{1}}} } \right)\sqrt {{\gamma _{2}}} }}{{{J^2} - {{\tilde E}_1}{{\tilde E}_2}}}} \right|^2}
\end{equation}
The second order correlation function at zero delay of the output field can be expressed in a compact form as a function of the coefficients $C_{jk}$ as
\begin{equation}\label{g2outS}
g_{\mathrm{out}}^{(2)}\left( 0 \right) = \frac{{\mathrm{Tr}\left( {c_{{\mathrm{out}}}^\dag c_{{\mathrm{out}}}^\dag {c_{{\mathrm{out}}}}{c_{{\mathrm{out}}}}\rho } \right)}}{{N_{{\mathrm{out}}}^2}} \simeq \frac{{{{\left| {{\gamma _1}{C_{20}} + {\gamma _2}{C_{02}}} \right|}^2} + {{\left| {{\gamma _1}{C_{20}} + \sqrt {{\gamma _1}{\gamma _2}} {C_{11}}} \right|}^2} + {{\left| {{\gamma _2}{C_{02}} + \sqrt {{\gamma _1}{\gamma _2}} {C_{11}}} \right|}^2}}}{{{{\left| {\sqrt {{\gamma _1}} {C_{10}} + \sqrt {{\gamma _2}} {C_{01}}} \right|}^4}}}
\end{equation}
\end{widetext}
We compare in Fig.A1 the analytical expression (\ref{g2out}) (see lower panels) to direct numerical solutions of the density matrix dynamics from Eq.(2) (see upper panels). We get a perfect agreement between the two for weak pump intensities.

Finally, we show that perfect output antibunching, namely $g_{\mathrm{out}}^{(2)}(0)=0$, cannot be obtained, differently from the intracavity field case. In particular, in order to make the numerator of expression (\ref{g2outS}) vanish, one would need
\begin{eqnarray}
\label{C02C20}
{{\mathrm{C}}_{02}} &=&  - \frac{{{\gamma _1}}}{{{\gamma _2}}}{{\mathrm{C}}_{20}}\\
\label{C11C20}
{{\mathrm{C}}_{11}} &=&  - \sqrt {\frac{{{\gamma _1}}}{{{\gamma _2}}}} {{\mathrm{C}}_{20}}\\
\label{C11C02}
{{\mathrm{C}}_{11}} &=&  - \sqrt {\frac{{{\gamma _2}}}{{{\gamma _1}}}} {{\mathrm{C}}_{02}} = \sqrt {\frac{{{\gamma _1}}}{{{\gamma _2}}}} {{\mathrm{C}}_{20}}
\end{eqnarray}
Obviously, these conditions can only be fulfilled by setting $C_{20}=C_{02}=C_{11}=0$, namely only in the unphysical situation in which the two-photon manifold of the Hilbert space is totally unoccupied. This remark is scarcely relevant to our main conclusions. In fact, the minimal values reached by the zero-delay two-photon correlation $g_{\mathrm{out}}^{(2)}(0)=0$, as computed from the full master equation, are to all practical purposes very small, and they are mostly determined by few-photon terms, thus beyond the two-photon limit assumed in the approximate analytical treatment above.

\bibliography{Input-Output}

\begin{thebibliography}{28}%
\makeatletter
\providecommand \@ifxundefined [1]{%
 \@ifx{#1\undefined}
}%
\providecommand \@ifnum [1]{%
 \ifnum #1\expandafter \@firstoftwo
 \else \expandafter \@secondoftwo
 \fi
}%
\providecommand \@ifx [1]{%
 \ifx #1\expandafter \@firstoftwo
 \else \expandafter \@secondoftwo
 \fi
}%
\providecommand \natexlab [1]{#1}%
\providecommand \enquote  [1]{``#1''}%
\providecommand \bibnamefont  [1]{#1}%
\providecommand \bibfnamefont [1]{#1}%
\providecommand \citenamefont [1]{#1}%
\providecommand \href@noop [0]{\@secondoftwo}%
\providecommand \href [0]{\begingroup \@sanitize@url \@href}%
\providecommand \@href[1]{\@@startlink{#1}\@@href}%
\providecommand \@@href[1]{\endgroup#1\@@endlink}%
\providecommand \@sanitize@url [0]{\catcode `\\12\catcode `\$12\catcode
  `\&12\catcode `\#12\catcode `\^12\catcode `\_12\catcode `\%12\relax}%
\providecommand \@@startlink[1]{}%
\providecommand \@@endlink[0]{}%
\providecommand \url  [0]{\begingroup\@sanitize@url \@url }%
\providecommand \@url [1]{\endgroup\@href {#1}{\urlprefix }}%
\providecommand \urlprefix  [0]{URL }%
\providecommand \Eprint [0]{\href }%
\providecommand \doibase [0]{http://dx.doi.org/}%
\providecommand \selectlanguage [0]{\@gobble}%
\providecommand \bibinfo  [0]{\@secondoftwo}%
\providecommand \bibfield  [0]{\@secondoftwo}%
\providecommand \translation [1]{[#1]}%
\providecommand \BibitemOpen [0]{}%
\providecommand \bibitemStop [0]{}%
\providecommand \bibitemNoStop [0]{.\EOS\space}%
\providecommand \EOS [0]{\spacefactor3000\relax}%
\providecommand \BibitemShut  [1]{\csname bibitem#1\endcsname}%
\let\auto@bib@innerbib\@empty
\bibitem [{\citenamefont {Kimble}(2008)}]{Kimble2008}%
  \BibitemOpen
  \bibfield  {author} {\bibinfo {author} {\bibfnamefont {H.~J.}\ \bibnamefont
  {Kimble}},\ }\href {http://dx.doi.org/10.1038/nature07127} {\bibfield
  {journal} {\bibinfo  {journal} {Nature}\ }\textbf {\bibinfo {volume} {453}},\
  \bibinfo {pages} {1023} (\bibinfo {year} {2008})}\BibitemShut {NoStop}%
\bibitem [{\citenamefont {Mabuchi}\ and\ \citenamefont
  {Doherty}(2002)}]{Mabuchi2002}%
  \BibitemOpen
  \bibfield  {author} {\bibinfo {author} {\bibfnamefont {H.}~\bibnamefont
  {Mabuchi}}\ and\ \bibinfo {author} {\bibfnamefont {A.~C.}\ \bibnamefont
  {Doherty}},\ }\href {\doibase 10.1126/science.1078446} {\bibfield  {journal}
  {\bibinfo  {journal} {Science}\ }\textbf {\bibinfo {volume} {298}},\ \bibinfo
  {pages} {1372} (\bibinfo {year} {2002})},\ \Eprint
  {http://arxiv.org/abs/http://www.sciencemag.org/content/298/5597/1372.full.pdf}
  {http://www.sciencemag.org/content/298/5597/1372.full.pdf} \BibitemShut
  {NoStop}%
\bibitem [{\citenamefont {{Carmichael}}\ \emph {et~al.}(1991)\citenamefont
  {{Carmichael}}, \citenamefont {{Brecha}},\ and\ \citenamefont
  {{Rice}}}]{Carmichael1991}%
  \BibitemOpen
  \bibfield  {author} {\bibinfo {author} {\bibfnamefont {H.~J.}\ \bibnamefont
  {{Carmichael}}}, \bibinfo {author} {\bibfnamefont {R.~J.}\ \bibnamefont
  {{Brecha}}}, \ and\ \bibinfo {author} {\bibfnamefont {P.~R.}\ \bibnamefont
  {{Rice}}},\ }\href {\doibase 10.1016/0030-4018(91)90194-I} {\bibfield
  {journal} {\bibinfo  {journal} {Optics Communications}\ }\textbf {\bibinfo
  {volume} {82}},\ \bibinfo {pages} {73} (\bibinfo {year} {1991})}\BibitemShut
  {NoStop}%
\bibitem [{\citenamefont {Birnbaum}\ \emph {et~al.}(2005)\citenamefont
  {Birnbaum}, \citenamefont {Boca}, \citenamefont {Miller}, \citenamefont
  {Boozer}, \citenamefont {Northup},\ and\ \citenamefont
  {Kimble}}]{Birnbaum2005}%
  \BibitemOpen
  \bibfield  {author} {\bibinfo {author} {\bibfnamefont {K.~M.}\ \bibnamefont
  {Birnbaum}}, \bibinfo {author} {\bibfnamefont {A.}~\bibnamefont {Boca}},
  \bibinfo {author} {\bibfnamefont {R.}~\bibnamefont {Miller}}, \bibinfo
  {author} {\bibfnamefont {A.~D.}\ \bibnamefont {Boozer}}, \bibinfo {author}
  {\bibfnamefont {T.~E.}\ \bibnamefont {Northup}}, \ and\ \bibinfo {author}
  {\bibfnamefont {H.~J.}\ \bibnamefont {Kimble}},\ }\href
  {http://dx.doi.org/10.1038/nature03804} {\bibfield  {journal} {\bibinfo
  {journal} {Nature}\ }\textbf {\bibinfo {volume} {436}},\ \bibinfo {pages}
  {87} (\bibinfo {year} {2005})}\BibitemShut {NoStop}%
\bibitem [{\citenamefont {He}\ \emph {et~al.}(2013)\citenamefont {He},
  \citenamefont {He}, \citenamefont {Wei}, \citenamefont {Wu}, \citenamefont
  {Atature}, \citenamefont {Schneider}, \citenamefont {Hofling}, \citenamefont
  {Kamp}, \citenamefont {Lu},\ and\ \citenamefont {Pan}}]{He2013}%
  \BibitemOpen
  \bibfield  {author} {\bibinfo {author} {\bibfnamefont {Y.-M.}\ \bibnamefont
  {He}}, \bibinfo {author} {\bibfnamefont {Y.}~\bibnamefont {He}}, \bibinfo
  {author} {\bibfnamefont {Y.-J.}\ \bibnamefont {Wei}}, \bibinfo {author}
  {\bibfnamefont {D.}~\bibnamefont {Wu}}, \bibinfo {author} {\bibfnamefont
  {M.}~\bibnamefont {Atature}}, \bibinfo {author} {\bibfnamefont
  {C.}~\bibnamefont {Schneider}}, \bibinfo {author} {\bibfnamefont
  {S.}~\bibnamefont {Hofling}}, \bibinfo {author} {\bibfnamefont
  {M.}~\bibnamefont {Kamp}}, \bibinfo {author} {\bibfnamefont {C.-Y.}\
  \bibnamefont {Lu}}, \ and\ \bibinfo {author} {\bibfnamefont {J.-W.}\
  \bibnamefont {Pan}},\ }\href {http://dx.doi.org/10.1038/nnano.2012.262}
  {\bibfield  {journal} {\bibinfo  {journal} {Nat Nano}\ }\textbf {\bibinfo
  {volume} {8}},\ \bibinfo {pages} {213} (\bibinfo {year} {2013})}\BibitemShut
  {NoStop}%
\bibitem [{\citenamefont {Matthiesen}\ \emph {et~al.}(2012)\citenamefont
  {Matthiesen}, \citenamefont {Vamivakas},\ and\ \citenamefont
  {Atat\"ure}}]{Matthiesen2012}%
  \BibitemOpen
  \bibfield  {author} {\bibinfo {author} {\bibfnamefont {C.}~\bibnamefont
  {Matthiesen}}, \bibinfo {author} {\bibfnamefont {A.~N.}\ \bibnamefont
  {Vamivakas}}, \ and\ \bibinfo {author} {\bibfnamefont {M.}~\bibnamefont
  {Atat\"ure}},\ }\href {\doibase 10.1103/PhysRevLett.108.093602} {\bibfield
  {journal} {\bibinfo  {journal} {Phys. Rev. Lett.}\ }\textbf {\bibinfo
  {volume} {108}},\ \bibinfo {pages} {093602} (\bibinfo {year}
  {2012})}\BibitemShut {NoStop}%
\bibitem [{\citenamefont {Reinhard}\ \emph {et~al.}(2012)\citenamefont
  {Reinhard}, \citenamefont {Volz}, \citenamefont {Winger}, \citenamefont
  {Badolato}, \citenamefont {Hennessy}, \citenamefont {Hu},\ and\ \citenamefont
  {Imamoglu}}]{Reinhard2012}%
  \BibitemOpen
  \bibfield  {author} {\bibinfo {author} {\bibfnamefont {A.}~\bibnamefont
  {Reinhard}}, \bibinfo {author} {\bibfnamefont {T.}~\bibnamefont {Volz}},
  \bibinfo {author} {\bibfnamefont {M.}~\bibnamefont {Winger}}, \bibinfo
  {author} {\bibfnamefont {A.}~\bibnamefont {Badolato}}, \bibinfo {author}
  {\bibfnamefont {K.~J.}\ \bibnamefont {Hennessy}}, \bibinfo {author}
  {\bibfnamefont {E.~L.}\ \bibnamefont {Hu}}, \ and\ \bibinfo {author}
  {\bibfnamefont {A.}~\bibnamefont {Imamoglu}},\ }\href
  {http://dx.doi.org/10.1038/nphoton.2011.321} {\bibfield  {journal} {\bibinfo
  {journal} {Nat Photon}\ }\textbf {\bibinfo {volume} {6}},\ \bibinfo {pages}
  {93} (\bibinfo {year} {2012})}\BibitemShut {NoStop}%
\bibitem [{\citenamefont {Bozyigit}\ \emph {et~al.}(2011)\citenamefont
  {Bozyigit}, \citenamefont {Lang}, \citenamefont {Steffen}, \citenamefont
  {Fink}, \citenamefont {Eichler}, \citenamefont {Baur}, \citenamefont
  {Bianchetti}, \citenamefont {Leek}, \citenamefont {Filipp}, \citenamefont
  {da~Silva}, \citenamefont {Blais},\ and\ \citenamefont
  {Wallraff}}]{Bozyigit2011}%
  \BibitemOpen
  \bibfield  {author} {\bibinfo {author} {\bibfnamefont {D.}~\bibnamefont
  {Bozyigit}}, \bibinfo {author} {\bibfnamefont {C.}~\bibnamefont {Lang}},
  \bibinfo {author} {\bibfnamefont {L.}~\bibnamefont {Steffen}}, \bibinfo
  {author} {\bibfnamefont {J.~M.}\ \bibnamefont {Fink}}, \bibinfo {author}
  {\bibfnamefont {C.}~\bibnamefont {Eichler}}, \bibinfo {author} {\bibfnamefont
  {M.}~\bibnamefont {Baur}}, \bibinfo {author} {\bibfnamefont {R.}~\bibnamefont
  {Bianchetti}}, \bibinfo {author} {\bibfnamefont {P.~J.}\ \bibnamefont
  {Leek}}, \bibinfo {author} {\bibfnamefont {S.}~\bibnamefont {Filipp}},
  \bibinfo {author} {\bibfnamefont {M.~P.}\ \bibnamefont {da~Silva}}, \bibinfo
  {author} {\bibfnamefont {A.}~\bibnamefont {Blais}}, \ and\ \bibinfo {author}
  {\bibfnamefont {A.}~\bibnamefont {Wallraff}},\ }\href
  {http://dx.doi.org/10.1038/nphys1845} {\bibfield  {journal} {\bibinfo
  {journal} {Nat Phys}\ }\textbf {\bibinfo {volume} {7}},\ \bibinfo {pages}
  {154} (\bibinfo {year} {2011})}\BibitemShut {NoStop}%
\bibitem [{\citenamefont {Ludwig}\ \emph {et~al.}(2012)\citenamefont {Ludwig},
  \citenamefont {Safavi-Naeini}, \citenamefont {Painter},\ and\ \citenamefont
  {Marquardt}}]{Ludwig2012}%
  \BibitemOpen
  \bibfield  {author} {\bibinfo {author} {\bibfnamefont {M.}~\bibnamefont
  {Ludwig}}, \bibinfo {author} {\bibfnamefont {A.~H.}\ \bibnamefont
  {Safavi-Naeini}}, \bibinfo {author} {\bibfnamefont {O.}~\bibnamefont
  {Painter}}, \ and\ \bibinfo {author} {\bibfnamefont {F.}~\bibnamefont
  {Marquardt}},\ }\href {\doibase 10.1103/PhysRevLett.109.063601} {\bibfield
  {journal} {\bibinfo  {journal} {Phys. Rev. Lett.}\ }\textbf {\bibinfo
  {volume} {109}},\ \bibinfo {pages} {063601} (\bibinfo {year}
  {2012})}\BibitemShut {NoStop}%
\bibitem [{\citenamefont {Nunnenkamp}\ \emph {et~al.}(2011)\citenamefont
  {Nunnenkamp}, \citenamefont {B\o{}rkje},\ and\ \citenamefont
  {Girvin}}]{Nunnenkamp2011}%
  \BibitemOpen
  \bibfield  {author} {\bibinfo {author} {\bibfnamefont {A.}~\bibnamefont
  {Nunnenkamp}}, \bibinfo {author} {\bibfnamefont {K.}~\bibnamefont
  {B\o{}rkje}}, \ and\ \bibinfo {author} {\bibfnamefont {S.~M.}\ \bibnamefont
  {Girvin}},\ }\href {\doibase 10.1103/PhysRevLett.107.063602} {\bibfield
  {journal} {\bibinfo  {journal} {Phys. Rev. Lett.}\ }\textbf {\bibinfo
  {volume} {107}},\ \bibinfo {pages} {063602} (\bibinfo {year}
  {2011})}\BibitemShut {NoStop}%
\bibitem [{\citenamefont {Rabl}(2011)}]{Rabl2011}%
  \BibitemOpen
  \bibfield  {author} {\bibinfo {author} {\bibfnamefont {P.}~\bibnamefont
  {Rabl}},\ }\href {\doibase 10.1103/PhysRevLett.107.063601} {\bibfield
  {journal} {\bibinfo  {journal} {Phys. Rev. Lett.}\ }\textbf {\bibinfo
  {volume} {107}},\ \bibinfo {pages} {063601} (\bibinfo {year}
  {2011})}\BibitemShut {NoStop}%
\bibitem [{\citenamefont {Stannigel}\ \emph {et~al.}(2012)\citenamefont
  {Stannigel}, \citenamefont {Komar}, \citenamefont {Habraken}, \citenamefont
  {Bennett}, \citenamefont {Lukin}, \citenamefont {Zoller},\ and\ \citenamefont
  {Rabl}}]{Stannigel2012}%
  \BibitemOpen
  \bibfield  {author} {\bibinfo {author} {\bibfnamefont {K.}~\bibnamefont
  {Stannigel}}, \bibinfo {author} {\bibfnamefont {P.}~\bibnamefont {Komar}},
  \bibinfo {author} {\bibfnamefont {S.~J.~M.}\ \bibnamefont {Habraken}},
  \bibinfo {author} {\bibfnamefont {S.~D.}\ \bibnamefont {Bennett}}, \bibinfo
  {author} {\bibfnamefont {M.~D.}\ \bibnamefont {Lukin}}, \bibinfo {author}
  {\bibfnamefont {P.}~\bibnamefont {Zoller}}, \ and\ \bibinfo {author}
  {\bibfnamefont {P.}~\bibnamefont {Rabl}},\ }\href {\doibase
  10.1103/PhysRevLett.109.013603} {\bibfield  {journal} {\bibinfo  {journal}
  {Phys. Rev. Lett.}\ }\textbf {\bibinfo {volume} {109}},\ \bibinfo {pages}
  {013603} (\bibinfo {year} {2012})}\BibitemShut {NoStop}%
\bibitem [{\citenamefont {Liew}\ and\ \citenamefont
  {Savona}(2010)}]{LiewPRL2010}%
  \BibitemOpen
  \bibfield  {author} {\bibinfo {author} {\bibfnamefont {T.~C.~H.}\
  \bibnamefont {Liew}}\ and\ \bibinfo {author} {\bibfnamefont {V.}~\bibnamefont
  {Savona}},\ }\href {\doibase 10.1103/PhysRevLett.104.183601} {\bibfield
  {journal} {\bibinfo  {journal} {Phys. Rev. Lett.}\ }\textbf {\bibinfo
  {volume} {104}},\ \bibinfo {pages} {183601} (\bibinfo {year}
  {2010})}\BibitemShut {NoStop}%
\bibitem [{\citenamefont {Carusotto}\ and\ \citenamefont
  {Ciuti}(2013)}]{Carusotto2013}%
  \BibitemOpen
  \bibfield  {author} {\bibinfo {author} {\bibfnamefont {I.}~\bibnamefont
  {Carusotto}}\ and\ \bibinfo {author} {\bibfnamefont {C.}~\bibnamefont
  {Ciuti}},\ }\href {\doibase 10.1103/RevModPhys.85.299} {\bibfield  {journal}
  {\bibinfo  {journal} {Rev. Mod. Phys.}\ }\textbf {\bibinfo {volume} {85}},\
  \bibinfo {pages} {299} (\bibinfo {year} {2013})}\BibitemShut {NoStop}%
\bibitem [{\citenamefont {Bamba}\ \emph {et~al.}(2011)\citenamefont {Bamba},
  \citenamefont {Imamo\ifmmode~\breve{g}\else \u{g}\fi{}lu}, \citenamefont
  {Carusotto},\ and\ \citenamefont {Ciuti}}]{BambaPRA2011}%
  \BibitemOpen
  \bibfield  {author} {\bibinfo {author} {\bibfnamefont {M.}~\bibnamefont
  {Bamba}}, \bibinfo {author} {\bibfnamefont {A.}~\bibnamefont
  {Imamo\ifmmode~\breve{g}\else \u{g}\fi{}lu}}, \bibinfo {author}
  {\bibfnamefont {I.}~\bibnamefont {Carusotto}}, \ and\ \bibinfo {author}
  {\bibfnamefont {C.}~\bibnamefont {Ciuti}},\ }\href {\doibase
  10.1103/PhysRevA.83.021802} {\bibfield  {journal} {\bibinfo  {journal} {Phys.
  Rev. A}\ }\textbf {\bibinfo {volume} {83}},\ \bibinfo {pages} {021802}
  (\bibinfo {year} {2011})}\BibitemShut {NoStop}%
\bibitem [{\citenamefont {Ferretti}\ \emph {et~al.}(2013)\citenamefont
  {Ferretti}, \citenamefont {Savona},\ and\ \citenamefont
  {Gerace}}]{FerrettiNJoP2013}%
  \BibitemOpen
  \bibfield  {author} {\bibinfo {author} {\bibfnamefont {S.}~\bibnamefont
  {Ferretti}}, \bibinfo {author} {\bibfnamefont {V.}~\bibnamefont {Savona}}, \
  and\ \bibinfo {author} {\bibfnamefont {D.}~\bibnamefont {Gerace}},\ }\href
  {http://stacks.iop.org/1367-2630/15/i=2/a=025012} {\bibfield  {journal}
  {\bibinfo  {journal} {New Journal of Physics}\ }\textbf {\bibinfo {volume}
  {15}},\ \bibinfo {pages} {025012} (\bibinfo {year} {2013})}\BibitemShut
  {NoStop}%
\bibitem [{\citenamefont {Bamba}\ and\ \citenamefont
  {Ciuti}(2011)}]{Bamba2011}%
  \BibitemOpen
  \bibfield  {author} {\bibinfo {author} {\bibfnamefont {M.}~\bibnamefont
  {Bamba}}\ and\ \bibinfo {author} {\bibfnamefont {C.}~\bibnamefont {Ciuti}},\
  }\href {\doibase 10.1063/1.3656250} {\bibfield  {journal} {\bibinfo
  {journal} {Applied Physics Letters}\ }\textbf {\bibinfo {volume} {99}},\
  \bibinfo {eid} {171111} (\bibinfo {year} {2011})}\BibitemShut {NoStop}%
\bibitem [{\citenamefont {Savona}(2013)}]{Savona2013}%
  \BibitemOpen
  \bibfield  {author} {\bibinfo {author} {\bibfnamefont {V.}~\bibnamefont
  {Savona}},\ }\href@noop {} {\bibfield  {journal} {\bibinfo  {journal}
  {arXiv}\ } (\bibinfo {year} {2013})},\ \Eprint
  {http://arxiv.org/abs/1302.5937} {1302.5937} \BibitemShut {NoStop}%
\bibitem [{\citenamefont {Xu}\ and\ \citenamefont {Li}(2013)}]{Xu2013}%
  \BibitemOpen
  \bibfield  {author} {\bibinfo {author} {\bibfnamefont {X.-W.}\ \bibnamefont
  {Xu}}\ and\ \bibinfo {author} {\bibfnamefont {Y.-J.}\ \bibnamefont {Li}},\
  }\href {http://stacks.iop.org/0953-4075/46/i=3/a=035502} {\bibfield
  {journal} {\bibinfo  {journal} {Journal of Physics B: Atomic, Molecular and
  Optical Physics}\ }\textbf {\bibinfo {volume} {46}},\ \bibinfo {pages}
  {035502} (\bibinfo {year} {2013})}\BibitemShut {NoStop}%
\bibitem [{\citenamefont {Majumdar}\ \emph {et~al.}(2012)\citenamefont
  {Majumdar}, \citenamefont {Bajcsy}, \citenamefont {Rundquist},\ and\
  \citenamefont {Vu\ifmmode \check{c}\else
  \v{c}\fi{}kovi\ifmmode~\acute{c}\else \'{c}\fi{}}}]{Majumdar2012}%
  \BibitemOpen
  \bibfield  {author} {\bibinfo {author} {\bibfnamefont {A.}~\bibnamefont
  {Majumdar}}, \bibinfo {author} {\bibfnamefont {M.}~\bibnamefont {Bajcsy}},
  \bibinfo {author} {\bibfnamefont {A.}~\bibnamefont {Rundquist}}, \ and\
  \bibinfo {author} {\bibfnamefont {J.}~\bibnamefont {Vu\ifmmode \check{c}\else
  \v{c}\fi{}kovi\ifmmode~\acute{c}\else \'{c}\fi{}}},\ }\href {\doibase
  10.1103/PhysRevLett.108.183601} {\bibfield  {journal} {\bibinfo  {journal}
  {Phys. Rev. Lett.}\ }\textbf {\bibinfo {volume} {108}},\ \bibinfo {pages}
  {183601} (\bibinfo {year} {2012})}\BibitemShut {NoStop}%
\bibitem [{\citenamefont {Ferretti}\ and\ \citenamefont
  {Gerace}(2012)}]{Ferretti2012}%
  \BibitemOpen
  \bibfield  {author} {\bibinfo {author} {\bibfnamefont {S.}~\bibnamefont
  {Ferretti}}\ and\ \bibinfo {author} {\bibfnamefont {D.}~\bibnamefont
  {Gerace}},\ }\href {\doibase 10.1103/PhysRevB.85.033303} {\bibfield
  {journal} {\bibinfo  {journal} {Phys. Rev. B}\ }\textbf {\bibinfo {volume}
  {85}},\ \bibinfo {pages} {033303} (\bibinfo {year} {2012})}\BibitemShut
  {NoStop}%
\bibitem [{\citenamefont {Lang}\ \emph {et~al.}(2011)\citenamefont {Lang},
  \citenamefont {Bozyigit}, \citenamefont {Eichler}, \citenamefont {Steffen},
  \citenamefont {Fink}, \citenamefont {Abdumalikov}, \citenamefont {Baur},
  \citenamefont {Filipp}, \citenamefont {da~Silva}, \citenamefont {Blais},\
  and\ \citenamefont {Wallraff}}]{Lang2011}%
  \BibitemOpen
  \bibfield  {author} {\bibinfo {author} {\bibfnamefont {C.}~\bibnamefont
  {Lang}}, \bibinfo {author} {\bibfnamefont {D.}~\bibnamefont {Bozyigit}},
  \bibinfo {author} {\bibfnamefont {C.}~\bibnamefont {Eichler}}, \bibinfo
  {author} {\bibfnamefont {L.}~\bibnamefont {Steffen}}, \bibinfo {author}
  {\bibfnamefont {J.~M.}\ \bibnamefont {Fink}}, \bibinfo {author}
  {\bibfnamefont {A.~A.}\ \bibnamefont {Abdumalikov}}, \bibinfo {author}
  {\bibfnamefont {M.}~\bibnamefont {Baur}}, \bibinfo {author} {\bibfnamefont
  {S.}~\bibnamefont {Filipp}}, \bibinfo {author} {\bibfnamefont {M.~P.}\
  \bibnamefont {da~Silva}}, \bibinfo {author} {\bibfnamefont {A.}~\bibnamefont
  {Blais}}, \ and\ \bibinfo {author} {\bibfnamefont {A.}~\bibnamefont
  {Wallraff}},\ }\href {\doibase 10.1103/PhysRevLett.106.243601} {\bibfield
  {journal} {\bibinfo  {journal} {Phys. Rev. Lett.}\ }\textbf {\bibinfo
  {volume} {106}},\ \bibinfo {pages} {243601} (\bibinfo {year}
  {2011})}\BibitemShut {NoStop}%
\bibitem [{\citenamefont {Boissonneault}\ \emph {et~al.}(2009)\citenamefont
  {Boissonneault}, \citenamefont {Gambetta},\ and\ \citenamefont
  {Blais}}]{Boissonneault2009}%
  \BibitemOpen
  \bibfield  {author} {\bibinfo {author} {\bibfnamefont {M.}~\bibnamefont
  {Boissonneault}}, \bibinfo {author} {\bibfnamefont {J.~M.}\ \bibnamefont
  {Gambetta}}, \ and\ \bibinfo {author} {\bibfnamefont {A.}~\bibnamefont
  {Blais}},\ }\href {\doibase 10.1103/PhysRevA.79.013819} {\bibfield  {journal}
  {\bibinfo  {journal} {Phys. Rev. A}\ }\textbf {\bibinfo {volume} {79}},\
  \bibinfo {pages} {013819} (\bibinfo {year} {2009})}\BibitemShut {NoStop}%
\bibitem [{\citenamefont {Koch}\ and\ \citenamefont {Le~Hur}(2009)}]{Koch2009}%
  \BibitemOpen
  \bibfield  {author} {\bibinfo {author} {\bibfnamefont {J.}~\bibnamefont
  {Koch}}\ and\ \bibinfo {author} {\bibfnamefont {K.}~\bibnamefont {Le~Hur}},\
  }\href {\doibase 10.1103/PhysRevA.80.023811} {\bibfield  {journal} {\bibinfo
  {journal} {Phys. Rev. A}\ }\textbf {\bibinfo {volume} {80}},\ \bibinfo
  {pages} {023811} (\bibinfo {year} {2009})}\BibitemShut {NoStop}%
\bibitem [{\citenamefont {Aldana}\ \emph {et~al.}(2013)\citenamefont {Aldana},
  \citenamefont {Bruder},\ and\ \citenamefont {Nunnenkamp}}]{Aldana2013}%
  \BibitemOpen
  \bibfield  {author} {\bibinfo {author} {\bibfnamefont {S.}~\bibnamefont
  {Aldana}}, \bibinfo {author} {\bibfnamefont {C.}~\bibnamefont {Bruder}}, \
  and\ \bibinfo {author} {\bibfnamefont {A.}~\bibnamefont {Nunnenkamp}},\
  }\href@noop {} {\enquote {\bibinfo {title} {On the equivalence between an
  optomechanical system and a kerr medium},}\ } (\bibinfo {year} {2013}),\
  \Eprint {http://arxiv.org/abs/arXiv:1306.0415} {arXiv:1306.0415} \BibitemShut
  {NoStop}%
\bibitem [{\citenamefont {Collett}\ and\ \citenamefont
  {Gardiner}(1984)}]{CollettPRA1984}%
  \BibitemOpen
  \bibfield  {author} {\bibinfo {author} {\bibfnamefont {M.~J.}\ \bibnamefont
  {Collett}}\ and\ \bibinfo {author} {\bibfnamefont {C.~W.}\ \bibnamefont
  {Gardiner}},\ }\href {\doibase 10.1103/PhysRevA.30.1386} {\bibfield
  {journal} {\bibinfo  {journal} {Physical Review A}\ }\textbf {\bibinfo
  {volume} {30}},\ \bibinfo {pages} {1386} (\bibinfo {year}
  {1984})}\BibitemShut {NoStop}%
\bibitem [{\citenamefont {Gardiner}\ and\ \citenamefont
  {Collett}(1985)}]{GardinerPRA1985}%
  \BibitemOpen
  \bibfield  {author} {\bibinfo {author} {\bibfnamefont {C.~W.}\ \bibnamefont
  {Gardiner}}\ and\ \bibinfo {author} {\bibfnamefont {M.~J.}\ \bibnamefont
  {Collett}},\ }\href {\doibase 10.1103/PhysRevA.31.3761} {\bibfield  {journal}
  {\bibinfo  {journal} {Phys. Rev. A}\ }\textbf {\bibinfo {volume} {31}},\
  \bibinfo {pages} {3761} (\bibinfo {year} {1985})}\BibitemShut {NoStop}%
\bibitem [{\citenamefont {Qu}\ \emph {et~al.}(1992)\citenamefont {Qu},
  \citenamefont {Xiao}, \citenamefont {Holliday}, \citenamefont {Singh},\ and\
  \citenamefont {Kimble}}]{Qu1992}%
  \BibitemOpen
  \bibfield  {author} {\bibinfo {author} {\bibfnamefont {Y.}~\bibnamefont
  {Qu}}, \bibinfo {author} {\bibfnamefont {M.}~\bibnamefont {Xiao}}, \bibinfo
  {author} {\bibfnamefont {G.~S.}\ \bibnamefont {Holliday}}, \bibinfo {author}
  {\bibfnamefont {S.}~\bibnamefont {Singh}}, \ and\ \bibinfo {author}
  {\bibfnamefont {H.~J.}\ \bibnamefont {Kimble}},\ }\href {\doibase
  10.1103/PhysRevA.45.4932} {\bibfield  {journal} {\bibinfo  {journal} {Phys.
  Rev. A}\ }\textbf {\bibinfo {volume} {45}},\ \bibinfo {pages} {4932}
  (\bibinfo {year} {1992})}\BibitemShut {NoStop}%
\end{thebibliography}%

\end{document}